\documentclass[journal=jpccck]{achemso}
\usepackage[version=3]{mhchem} 
\usepackage[T1]{fontenc}       
\usepackage[usenames, dvipsnames]{color}
\usepackage{longtable}
\usepackage{multirow}
\usepackage{makecell}

\title{Orbital-hybridization-created optical excitations in Li\textbf{$_2$}GeO\textbf{$_{3}$}}
\author{Vo Khuong Dien}
\email{vokhuongdien@gmail.com}
\affiliation[National Cheng Kung University]{Department of Physics, National Cheng Kung University, Tainan, 70101, Taiwan}
\author{Hai Duong Pham}%
\affiliation[National Cheng Kung University]{Department of Physics, National Cheng Kung University, Tainan, 70101, Taiwan}
\author{Ngoc Thanh Thuy Tran}
\affiliation[National Cheng Kung University]{Hierarchical Green Energy Materials, Hi- esearch Center, National Cheng Kung University, Taiwan}
\author{Nguyen Thi Han}
\affiliation[National Cheng Kung University]{Department of Physics, National Cheng Kung University, Tainan, 70101, Taiwan}
\author{Thi My Duyen Huynh}
\affiliation[National Cheng Kung University]{Department of Physics, National Cheng Kung University, Tainan, 70101, Taiwan}
\author{Thi Dieu Hien Nguyen}
\email{nguyenhien1901@gmail.com}
\affiliation[National Cheng Kung University]{Department of Physics, National Cheng Kung University, Tainan, 70101, Taiwan}
\author{Ming Fa-Lin}
\email{mflin@mail.ncku.edu.tw}
\affiliation[National Cheng Kung University]{Department of Physics, National Cheng Kung University, Tainan, 70101, Taiwan}
\affiliation[National Cheng Kung University]{Hierarchical Green Energy Materials, Hi- esearch Center, National Cheng Kung University, Taiwan}
\begin{document}
\begin{abstract}
Li$_2$GeO$_{3}$, a ternary electrolyte compound of Li$^+$-based battery, presents the unusual essential properties. The main features are thoroughly explored from the first-principles calculations. The concise pictures, the critical orbital hybridizations in Li-O and Ge-O bonds, are clearly examined through the optimal {$Moir\acute{e}$} superlattice, the atom-dominated electronic energy spectrum, the spatial charge densities, the atom- and orbital-decomposed van Hove singularities, and the strong optical responses.  The unusual optical transitions cover the red-shift optical gap, 16 frequency-dependent absorption structures and the most prominent plasmon mode in terms of the dielectric functions, energy loss functions, reflectance spectra, and absorption coefficients. Optical excitations, depending on the directions of electric polarization, are strongly affected by excitonic effects. The close combinations of electronic and optical properties can identify a significant orbital hybridization for each available excitation channel. The developed theoretical framework will be very useful in fully understanding the diverse phenomena of cathode/electrolyte/anode materials in Lithium ion-based batteries.
\end{abstract}

\maketitle


\section{INTRODUCTION}

\hspace*{1cm}Lithium-ion batteries (LIBs) dominate in commercial purposes due to their high-performance energy resources, e.g. high power, energy densities, and high reliability, as well as other certain fundamental merits including affordable price, long life cycle, and friendly environment \cite{1,2,3}. The recent experimental progress shows many applications of LIBs, such as electric vehicles (EV) and hybrid EV and mobile devices, which require the optimum in producing efficient energy. A LIB principally consists of a cathode (positive electrode), an anode (negative electrode), and an ionically conductive Li$^+$-containing electrolyte, in which the third component is closely related to the unusual transport of the positive lithium ions (Li$^+$) between two electrodes \cite{1,2,3}. Depending on the combined alternative materials of three components, LIBs can provide various performances, e.g, a specific energy density of 100 to 250 W.h/Kg \cite{4,5}, the volumetric energy density from 250 to 680 W.h/L \cite{6}, a specific power density in the range of 300 to 1500 W/kg \cite{7,8}, and a faster charging time (80 \% of charge of states in 15 mins) \cite{9}. However, recently used electrolytes in LIBs are not compatible with later developed high-voltage positive electrodes, which are one of the most effective ways of increasing the energy density.\\
\hspace*{1cm}Nowadays, solid electrolytes are rapidly emerging as promising alternatives given their wider electrochemical window of stability. As potential electrolyte candidates, Li-Ge-O compounds exhibit a large ionic conductivity (1.5$\times$10$^{-5}\Omega$.cm$^{-1}$ for Li$_2$GeO$_3$), in which electronic conductivity is negligible with a high ionic transference number \cite{10}. Furthermore, Li$_2$GeO$_3$ shows a wide cycling stability with a reserved charge capacity of 725 mAhg$^{-1}$ after 300 cycles at 50 mAg$^{-1}$\cite{11}. As for battery safety, Li$_2$GeO$_3$ is suitable for the selection of solid electrolytes according to the reasonable decreasing interface resistance. On the other hand, previous reports evidence that the Li$_2$GeO$_3$ sample also exhibit good optical properties such as the high transparency in the visible region, the strong Stock shift of the self-trap exciton \cite{12}, the polar orthorhombic symmetry \cite{13,14} suggests that this material is useful for piezoelectric, pyroelectric, and electrochromic energy storage devices.\\
\hspace*{1 cm}Up to now, a lot of high-resolution measurements on essential physical properties of anodes/cathodes/electrolytes in Li$^+$-based batteries, e.g, optimized geometric structures, band structures, optical properties and transport have been conducted. X-Ray diffraction \cite{11} and low-energy electron diffraction (LEED \cite{15}) are available for investigating the 3D lattice symmetries, especially Li-X-O related systems. Angle-resolved photoemission spectroscopy (ARPES \cite{16}) is a powerful experimental technique directly measuring the single particle spectral function, depending on wave vectors and frequency. Optical properties, such as energy loss functions, reflectance, absorptance, refractivity,and excitation \cite{17,18,19} could be determined through spectroscopy measurements. For example, analysis of the specular reflection from the polished surface of a solid can give the complex index of refraction using the Kramers-Kronig method.\\
\hspace*{1cm}As for Li-Ge-O compounds, a few delicate experimental measurements, such as X-Ray diffraction \cite{11} and photoluminescence measurement \cite{12} were carried out. Up to date, concise physical and chemical pictures that closely relate to the performance of an energy storage system have not been proposed for the essential properties of the Li-Ge-O compound. Furthermore, the theoretical electrolytes related to Li-Ge-O compounds with band structures, optical properties and other predictions about chemical/physical combinations are still absent. In this work, the close connection of the geometric structure, electronic and optical excitations of Li$_2$GeO$_3$ solid states electrolyte was identified for the first time.\\   
\hspace*{1cm}The theoretical framework, being based on the significant orbital hybridizations in chemical bonds \cite{20,21}, is developed by examining the essential properties in electrolyte materials of Li$^+$-related batteries \cite{22}. This strategy is based on the first-principles calculations on an optimal lattice symmetry with position-dependent chemical bondings, the atom-dominated band structure at different energy ranges, the spatial charge densities due to various orbitals, and the atom- and orbital-projected van Hove singularities related to orbital overlaps. The energy-decomposed single-/multi-orbital hybridizations will be utilized to account for the optical threshold frequency, a lot of prominent absorption structures, a very strong plasmon response in terms of the dielectric functions, energy loss functions, reflectance spectra, and absorption coefficients under the distinct electric polarizations. This is the first work that successfully combines the geometric, electronic and optical properties with the energy-dependent orbital hybridizations. Most predicted results require high-resolution experimental examinations.
\section{COMPUTATIONAL DETAILS}
\hspace*{1cm}We used the density functional theory (DFT) method via the Vienna Ab-initio Simulation Package (VASP) \cite{23} to perform the optimization of the structure and calculation the electronic and optical properties. The Perdew-Burke-Ernzerhof (PBE) generalized gradient approximation  was used for the exchange-correlation functional \cite{24}. The interaction between the ions and valence electrons was described by the projector augmented wave (PAW) method \cite{25}. The cutoff energy for the expansion of the plane wave basis was set to 500 eV. The Brillouin zone was integrated with a special k-point mesh of 15$\times$15$\times$15 in the Monkhorst-Pack sampling technique \cite{26} for geometric optimization. The convergence condition of the ground-state is set to be 10$^{-8}$ eV between two consecutive simulation steps, and all atoms were allowed to fully relax during geometric optimization until the Hellmann-Feynman force acting on each atom was smaller than 0.01 eV.\\
\hspace*{1 cm} Subsequent to the DFT results, the quasi-particle Green's function and the screened Coulomb interactions (G$_0$W$_0$) approach \cite{27} using 250 eV energy cutoff for the response functions and the Brillouin zone was integrated with a special k-points mesh of 8$\times$8$\times$8 in the Monkhorst-Pack sampling technique to obtain the corrected density of states and electronic band structure and quasi-particle excitations of Li$_2$GeO$_3$. Regarding the optical response beyond the independent particle approach, the electron-hole interaction was taken into account by solving the standard Bethe-Salpeter equation (BSE) within the Tamm-Dancoff approximation \cite{28}. The k-point sampling, energy cutoff, and number of bands setting the same as in the GW calculation. In this calculation, the 20 highest occupied valence bands (VBs) and 8 lowest unoccupied conduction bands (CBs) are included as a basis for the excitonic states with a photon energy region from 0 eV to 28 eV. In addition, the broadening parameter $\gamma$, which arises from various de-excitation mechanisms, was set at 0.1 eV to make the spectrum more accurate for the given k-point meshes. All parameters in this study have been checked for convergence of the calculations.
\section{RESULTS AND DISCUSSIONS}
\begin{figure*}[htb]
       \includegraphics[scale=0.4]{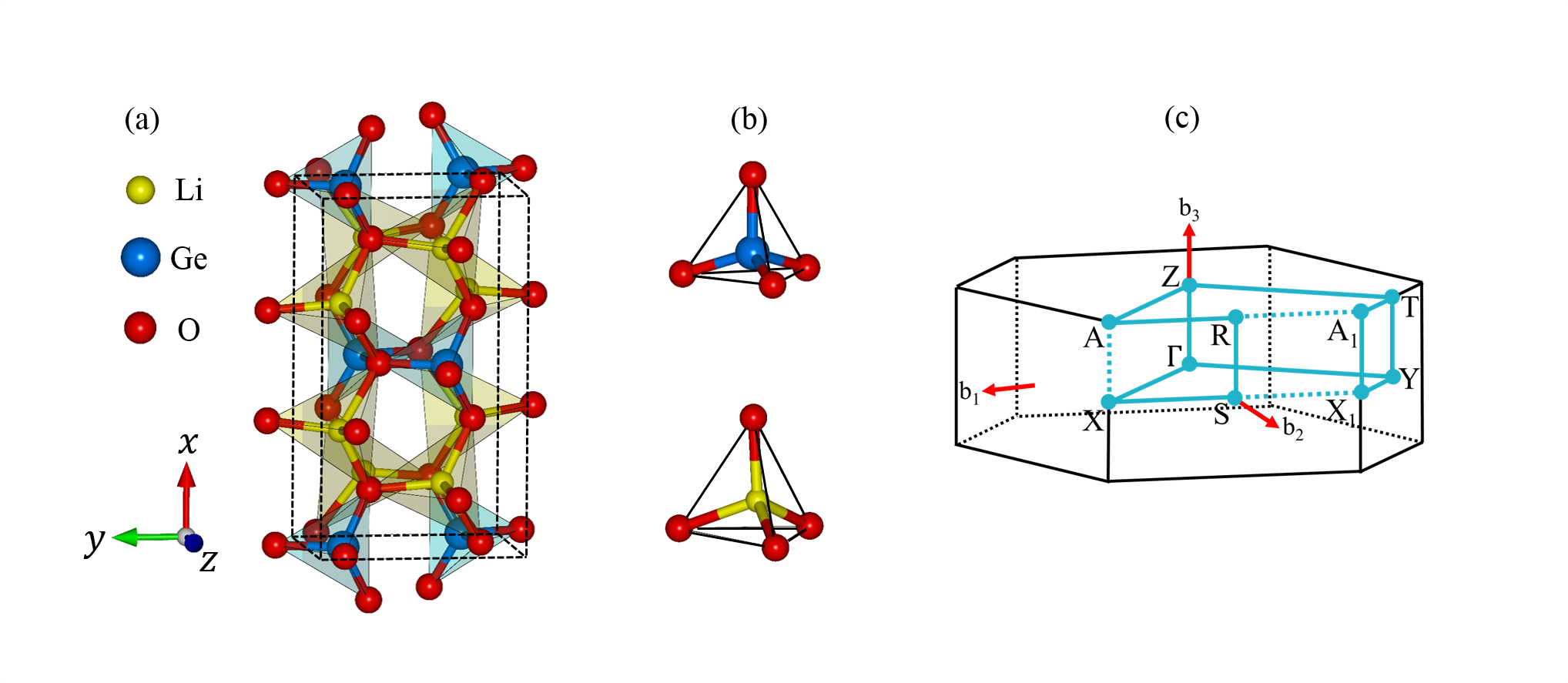}
       \caption{The optimal geometric structure of Li$_2$GeO$_3$ ternary compound, (b) oxygen atoms around each Li/Ge one, and (c) the first Brillouin zone.}
       \label{fgr:1}
\end{figure*}   
\hspace*{1 cm} The ternary compound, Li$_2$GeO$_3$ with 24 atoms within a conventional unit cell, is chosen for a model study in illustrating the geometric, electronic and optical properties. The optimal lattice, as clearly shown in Fig. 1, belongs to an orthorhombic structure. Each Li/Ge atom is surrounded by four O atoms in a tetrahedral form. There exist 32 Li-/16 Ge-O bonds, in which the former and the latter display large fluctuations about their lengths ($\sim$ 1.93-2.12 {\AA} and $\sim$ 1.72-1.84 {\AA}, respectively). The calculated lattice constants along the $x$-, $y$- and $z$-directions, (9.61, 5.46, 4.87 {\AA}), are very close to the X-ray diffraction measurements, (9.60, 5.50, 4.85 {\AA}) in Ref. [11] and (9.63, 5.48, 4.84 {\AA}) in Ref. [29]. Obviously, the greatly non-uniform chemical/physical environment indicates the importance of orbital hybridizations in chemical bondings. Furthermore, it might be very useful for supporting the ion transport in electrolyte materials. This behavior will be responsible for the unusual effects of {$Moir\acute{e}$} superlattice and the highly anisotropic optical transitions.\\
\begin{figure*}[htb]
	\includegraphics[scale=0.5]{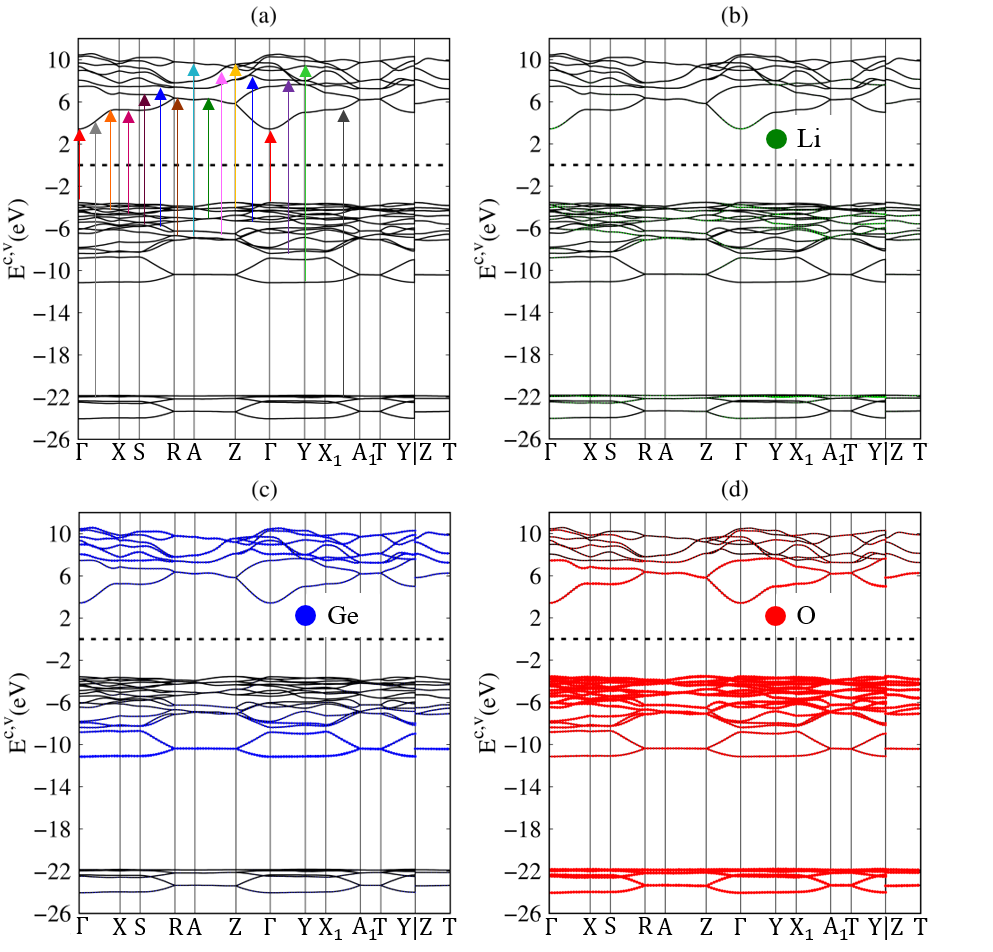}
	\label{fgr:2}
	\caption{(a) Band structure along the high-symmetry points in the wave-vector space, with (b) Li- (enlarged five times), (c) Ge- and (d) O-atom dominances (green, blue and red circles).}
\end{figure*}
\hspace*{1 cm}Li$_2$GeO$_3$ exhibits a rich and unique band structure. The main features in Fig. 2(a) show a lot of valence and conduction subbands in a wide energy range (-26 eV $<$ E$^{c,v} <$ 12 eV), the high asymmetry of hole and electron energy spectra about the Fermi level (E$_F =$ 0), an indirect band gap of E$_g^i=$ 6.9 eV related to the highest occupied state and the lowest unoccupied one, respectively, at the Z and $\Gamma$ points, the parabolic/oscillatory/linear/partially flat energy dispersions with various band-edge states, and the atom-dominated electronic states in the different energy regions. Very interestingly, such results are associated with many significant orbital hybridizations in Li-O and Ge-O bonds, being clearly identified from the close combinations of the atom- and/or orbital-dependent band structure (Figs. 2(b), 2(c) and 2(d)), charge density (Fig. 3) and van Hove singularities (Fig. 4).\\
\hspace*{1 cm}The electronic structures can be classified into five categories based on the various atom and orbital contributions, as clearly indicated in Figs. 2(b), 2(c) and 2(d) by the green, blue and red circles for Li, Ge and O, respectively. The (I), (II), (III), (IV) and (V) regions, respectively, correspond to E$^c >$ 7.6 eV, 3.2 eV $<$ E$^c <$ 7.6 eV, -6.0 eV $<$ E$^v <$ -3.8 eV, -10.6 eV $<$ E$^v <$ -6.0 eV, and -24.5 eV $<$ E$^v <$ -21.8 eV. It should be noted that valence subbands disappear between the (IV) and (V) regions.  The Li atoms only make very weak, yet still significant contributions within a whole range of band structures (small green circles in Fig. 2(b)), i.e., the unusual essential properties disappear in the absence of Li-O bonds. The energy-spectrum subgroups are qualitatively characterized by (I) (Ge, O) co-dominance, (II) (Ge, O) co-dominance, (III) O dominance, (IV) (Ge, O) co-dominance and (V) O dominance, being supported by the atom- and orbital-projected density of states (discussed later in Fig. 4). The specific orbital hybridizations in chemical bonds will be identified to be associated with the critical energy bands in revealing the prominent absorption structures.\\
\hspace*{1 cm}The spatial charge distributions before/after chemical bondings can provide some very useful information about the first-step orbital hybridizations in Li-O and Ge-O bonds. An isolated Li atom has an isotropic charge density (Fig. 3(a)), in which the inner and outer regions (the red and white parts) arise from 1s and 2s orbitals, respectively. The similar, but wider distribution, which corresponds to the O case (Fig. 3(b)), is associated with (1s, 2s) and (2p$_x$, 2p$_y$, 2p$_z$) orbitals. Furthermore, the highest charge density appears around Ge (Fig. 3(c)), with the separate ranges of (4s, those below it) and (4p$_x$, 4p$_y$, 4p$_z$) orbitals. As for the Li-O bonds, the outer/inner regions show the obvious/minor deformations along three electric-polarization directions (Figs. 3(d), 3(e) and 3(f)), especially for the neighboring ones. These clearly indicate the multi-orbital hybridizations of 2s-(2p$_x$, 2p$_y$, 2p$_z$)/the single-orbital hybridization of 2s-2s. In addition to the white regions (Figs. 3(g), 3(h) and 3(i)), the red ones near the Ge and O atoms present observable changes, suggesting the significant bondings of (4s, 4p$_x$, 4p$_y$, 4p$_z$)-(2s, 2p$_x$, 2p$_y$, 2p$_z$). While the above-mentioned analyses are combined with the atom- and orbital-projected van Hove singularities, the orbital-hybridization-dominated band-edge states will be identified. They are responsible for a lot of strong absorption structures (Figs. 5 and 6).\\           
\begin{figure}[htb]
	\includegraphics[scale=0.4]{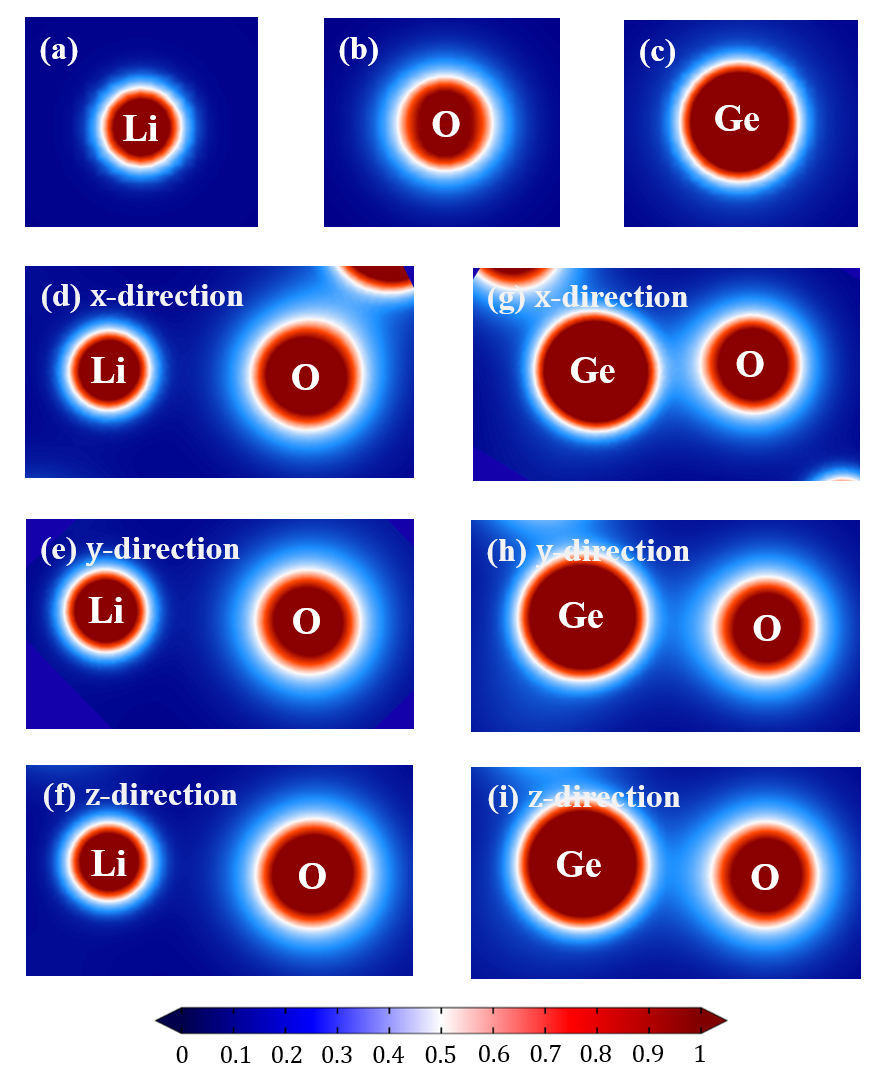}
	\label{fgr:3}
	\caption{Comparison of an isolated (a)/(b)/(c) Li/O/Ge atom, the spatial charge distributions related to the significant orbital hybridizations in (d)/(e)/(f) Li-O and (g)/(h)/(i) Ge-O bonds along the $x$-/$y$-/$z$-directions for the shortest chemical bonds. }
\end{figure}
\begin{figure}[htb]
	\includegraphics[scale=0.3]{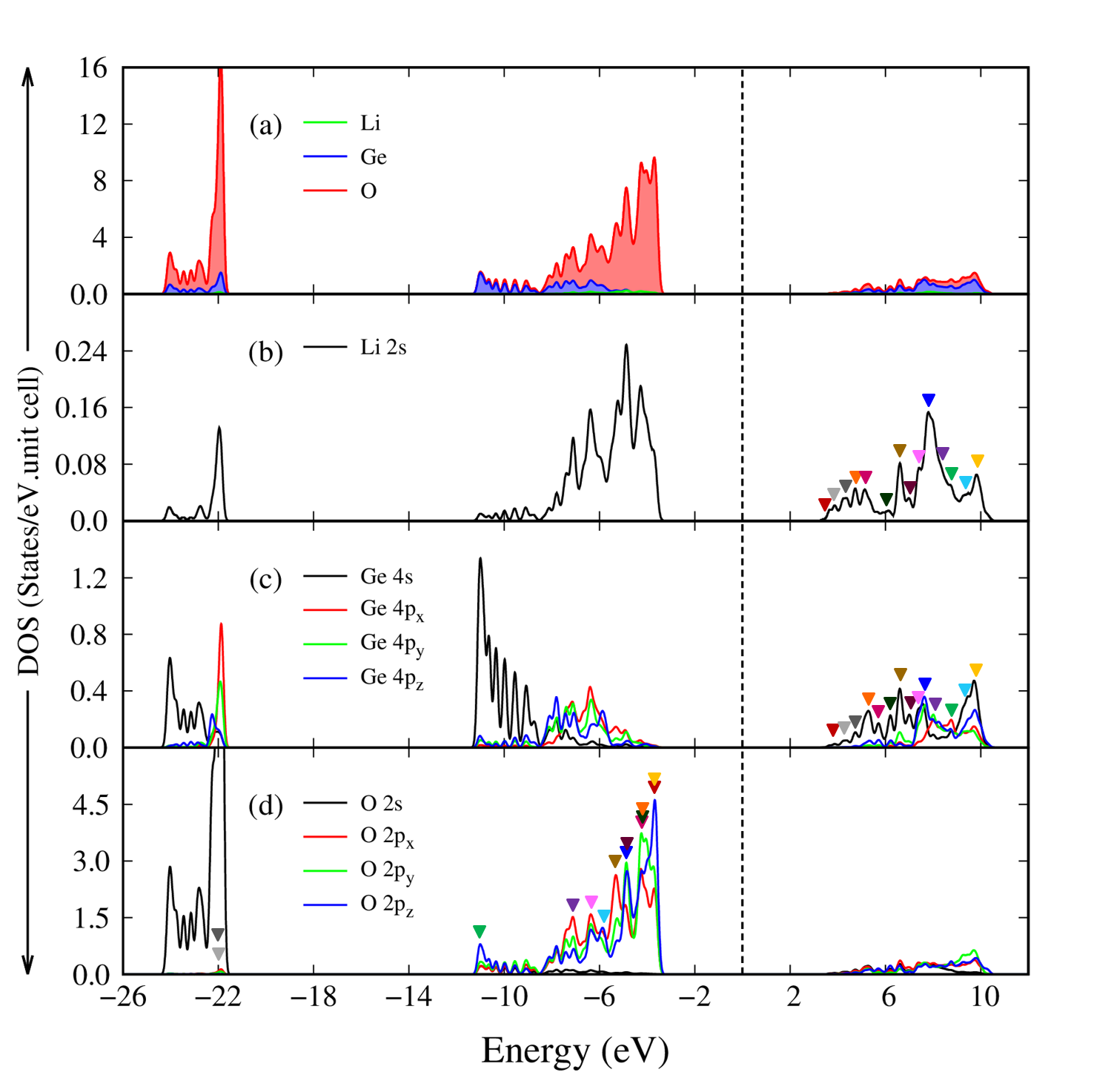}
	\label{fgr:4}
	\caption{Density of states for different components: (a) the total magnitude with Li-, Ge- and O-atom contributions, (b)  Li- and 2s-decomposed results, (c)  Ge- and (4s, 4p$_x$, 4p$_y$, 4p$_z$)-projected calculations, and (d) O- and (2s, 2p$_x$, 2p$_y$, 2p$_z$)-related ones. }
\end{figure}
\hspace*{1cm}The atom- and orbital-projected density of states (Fig. 4), being supported by the atom-dominated band structure (Fig. 2) and charge density distribution (Fig. 3), can clearly identify the energy-dependent orbital hybridizations. They come from the band-edge states of various energy dispersions, such as, the extreme, saddle and dispersionless ones. Both critical points in the energy-wave-vector spaces and dimensionality fully determine their special structures \cite{20,21}. As a result of too many subbands and broadening effects, Fig. 4(a) only displays the prominent asymmetric peaks and shoulders (the black curve). The van Hove singularities of different atoms/orbitals can merge together (Figs. 4(a)-4(d)), clearly indicating the specific orbital hybridizations. According to their strong co-relations, the five subgroups of band structure are further examined to be dominant through (I) 2s-(2s, 2p$_x$, 2p$_y$, 2p$_z$) \& (4s, 4p$_x$, 4p$_y$, 4p$_z$)-(2s, 2p$_x$, 2p$_y$, 2p$_z$), (II) 2s-(2p$_x$, 2p$_y$, 2p$_z$) \& 4s-(2p$_x$, 2p$_y$, 2p$_z$), (III) 2s-(2p$_x$, 2p$_y$, 2p$_z$) \& (4s, 4p$_x$, 4p$_y$, 4p$_z$)-(2p$_x$, 2p$_y$, 2p$_z$), (IV) 2s-(2p$_x$, 2p$_y$, 2p$_z$) \& 4s-(2p$_x$, 2p$_y$, 2p$_z$) and (V) 2s-2s \& 4s-2s. Such identifications of significant orbital bondings would become more delicate under strong optical responses, mainly owing to the greatly reduced energy width of the excitation frequency. \\
      \begin{figure*}[htb]
	\includegraphics[scale=0.35]{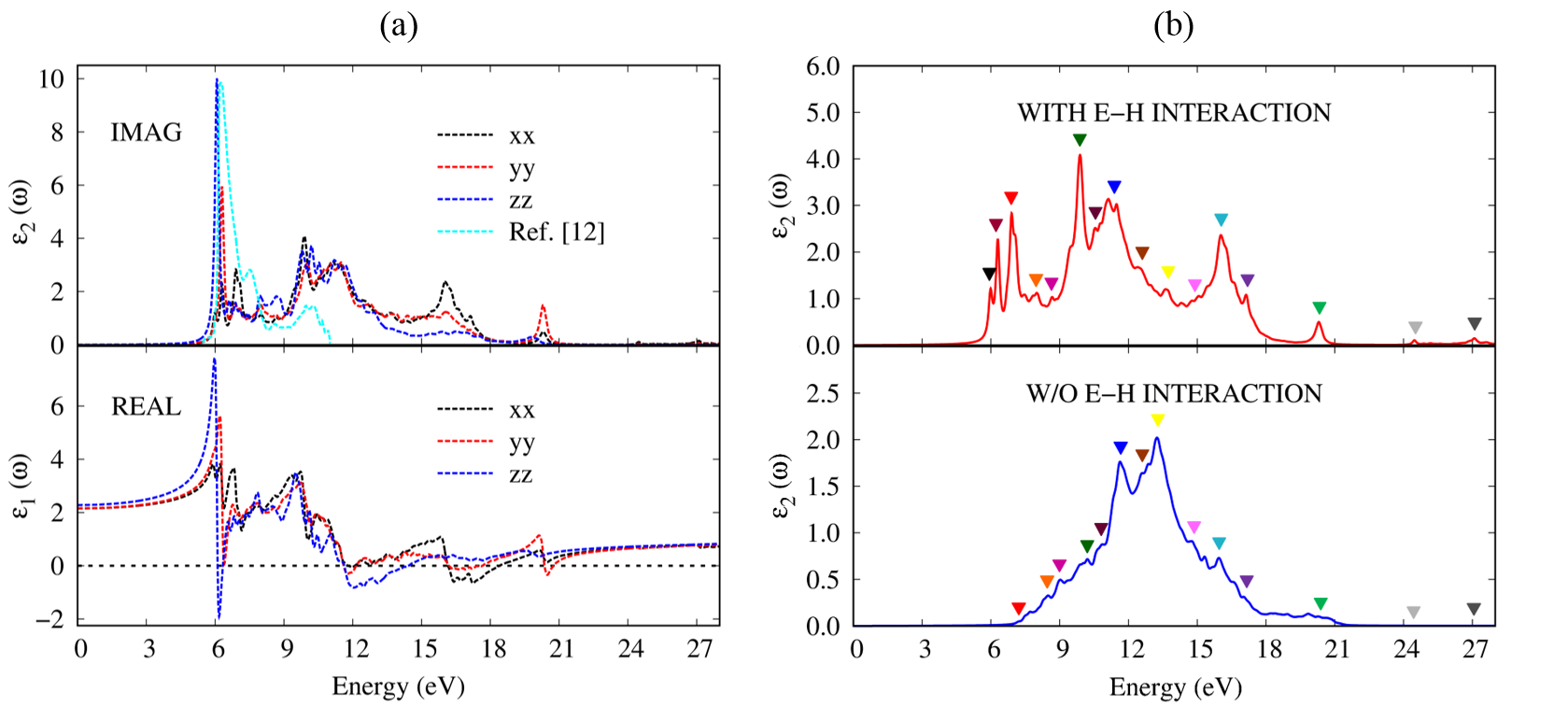}
	\label{fgr:5}
	\caption{(a) The imaginary- and real-parts of the dielectric functions with the excitonic effects under three electric-polarization directions [the black, red and blue curves]. (b) Comparison of the imaginary parts of the dielectric function with and without excitonic effects under xx-polarization direction.}
\end{figure*}
\hspace*{1cm}After the perturbation of an electromagnetic (EM) wave, electrons are vertically excited from the occupied to unoccupied states. Under the single-particle picture (the linear Kubo formula in Ref. [30]), the optical absorption spectrum is mainly determined by the joint density of states related to the valence and conduction subbands and the square of the electric dipole moment, respectively, corresponding to the number and intensity of available optical transitions. The excited valence holes and conduction electrons simultaneously come into existence during the optical excitations; furthermore, they have a tendency to combine through the attractive Coulomb potentials under the suitable condition, e.g., the large band gap for the suppression of temperature broadening. The coupled quasiparticles, the stable excitons, might strongly affect the main features of the optical absorption spectra since they make important contributions to many-body effects. This mean that they take part in the various-order excitation processes using the Coulomb scatterings, leading to the dramatic transformations of photons. The excitonic effects, which are closely related to the critical orbital hybridizations, are the focus of this study.\\
\hspace*{1cm}The imaginary and real parts of dielectric functions are very useful in understanding the available optical transitions. They can fully examine the orbital-related prominent responses by linking the atom-dominated band structure (Fig. 2), charge density (Fig. 3), and atom - $\&$ orbital-decomposed van Hove singularities (Fig. 4). The former ($\epsilon_2(\omega)$), as clearly shown in Fig. 5(a), presents the excitation characteristics of available channels in the absence of a screening effect of all valence charges. An optical gap (E$_g^o$), which corresponds to the threshold absorption frequency, is about 5.95/5.92/5.87 eV under the many-body effects for the $x$-/$y$-/$z$-directions of electric polarizations (the black/red/blue curves). However, it is $\sim$ 7.00 eV purely through the single-particle optical excitations for \textbf{E}$\|$$\hat{\textbf{x}}$ (the blue curve in Fig. 5(b)). Compared with band gap of E$_g^i$ = 6.9 eV, the great red shift clearly indicates the very strong Coulomb couplings between the excited holes and electrons. The composite quasiparticles should be quite stable, so that they are expected to survive at room temperature. The theoretical predictions are relatively easily examined by the various optical spectroscopies \cite{12, 17,18,19}.\\ 
\hspace*{1cm}Very interestingly, there exist 16 pronounced peak/shoulder absorptions (the distinct colored triangles in Figs. 5(b),S1 and S2). In which, the appearance of these certain optical-spectral structures is associated with the transitions from band-edge states/Van Hove singularities of electronic band structure/orbital-projected density of states. Furthermore, the formation of the stable excitonic states around the respective extrema of the band-edge state also creates extra prominence peaks below the band gap. However, it can not strongly modify intrinsic chemical bondings (the identity of the joined of density of states). Based on the close connection of the concise physical and chemical pictures in atom-dominated band structure (Fig. 2), atom-, orbital-decomposed density of states (Fig. 4), and strong optical responses due to band-edge states. The frequency-dependent vertical transitions and orbital hybridizations are identified (detail in Table 1). The corresponding excitation of channels are marked with arrow heads in Fig. 2(a) and the triangle in Fig. 4. Generally, these channels are major/minor contributions from the electron transition between the O-2p occupied states and the Ge-(4s,4p)/Li-2s unoccupied states. For example,the threshold structure is marked by a red triangle in Fig. 5(b), which is due to the excitation from the valence flat band to the conduction minimum extremes at the $\Gamma$ point. It is mainly/slightly dominated by the O-2p$_z$ valence states and the Ge-4s/Li-2s conduction ones, indicated by the red arrow head in Fig. 2(a) and the red triangle in Fig. 4, respectively. Using this strategy, the specified mechanisms never previously revealed in studies for the various absorption structures could be achieved. This viewpoint has been successfully generalized to anode and cathode materials of Li$^+$-based batteries.\\
\begin{table*}[t]
	\caption{Prominent absorption structures: frequencies, colored indicators and identified orbital hybridizations}
	\label{tab:my-table}
	\begin{tabular}{cccc}
		\hline\hline
		\multicolumn{2}{c}{Excitation frequency (eV)} &
		\multicolumn{1}{c}{Colors of arrows} &
		\multicolumn{1}{c}{\begin{tabular}[c]{@{}c@{}}Specific orbital hybridizations \\ in Ge-O bonds\end{tabular}} \\ \hline
		\multicolumn{1}{c}{\begin{tabular}[c]{@{}c@{}}With excitonic\\  effect\end{tabular}} &
		\multicolumn{1}{c}{\begin{tabular}[c]{@{}c@{}}Without excitonic\\  effect\end{tabular}} &
		\multicolumn{1}{c}{} &
		\multicolumn{1}{c}{} \\ \hline
		\multicolumn{1}{c}{6}    & \multicolumn{1}{c}{--}   & \multicolumn{1}{c}{Black}       & \multicolumn{1}{c}{--}                                          \\ 
		\multicolumn{1}{c}{6.3}  & \multicolumn{1}{c}{--}   & \multicolumn{1}{c}{Dark-red}        & \multicolumn{1}{c}{--}                                      \\ 
		\multicolumn{1}{c}{7}    & \multicolumn{1}{c}{7}  & \multicolumn{1}{c}{Red}         & \multicolumn{1}{c}{Ge(4s) - O(2p$_z$)}                            \\ 
		\multicolumn{1}{c}{8}    & \multicolumn{1}{c}{8}  & \multicolumn{1}{c}{Orange}      & \multicolumn{1}{c}{Ge(4s) - O(2p$_x$,2p$_y$,2p$_z$)}              \\ 
		\multicolumn{1}{c}{8.8}  & \multicolumn{1}{c}{8.9}  & \multicolumn{1}{c}{Purple}      & \multicolumn{1}{c}{Ge(4s) - O(2p$_x$,2p$_y$,2p$_z$)}            \\ 
		\multicolumn{1}{c}{9.9}  & \multicolumn{1}{c}{10}    & \multicolumn{1}{c}{Green}       & \multicolumn{1}{c}{Ge(4s) - O(2p$_x$,2p$_y$,2p$_z$)}                \\ 
		\multicolumn{1}{c}{10.6} & \multicolumn{1}{c}{10.6}  & \multicolumn{1}{c}{Maroon}        & \multicolumn{1}{c}{Ge(4s) - O(2p$_x$,2p$_y$, 2p$_z$)}          \\ 
		\multicolumn{1}{c}{11.3} & \multicolumn{1}{c}{11.4} & \multicolumn{1}{c}{Blue}       & \multicolumn{1}{c}{Ge(4s,4p$_x$,4p$_y$,4p$_z$) - O(4p$_x$,4p$_y$, 4p$_z$)}       \\ 
		\multicolumn{1}{c}{12.5} & \multicolumn{1}{c}{12.5} & \multicolumn{1}{c}{Brown}      & \multicolumn{1}{c}{Ge(4s) - O(2p$_x$,2p$_y$)}   \\ 
		\multicolumn{1}{c}{13.6} & \multicolumn{1}{c}{13.5} & \multicolumn{1}{c}{Yellow}        & \multicolumn{1}{c}{Ge(4s,4p$_x$,4p$_y$,4p$_z$) - O(2p$_x$,2p$_y$,2p$_z$)}  \\ 
		\multicolumn{1}{c}{14.9} & \multicolumn{1}{c}{14.9} & \multicolumn{1}{c}{Pink}        & \multicolumn{1}{c}{Ge(4s) - O(2p$_x$,2p$_y$,2p$_z$)}          \\ 
		\multicolumn{1}{c}{16}   & \multicolumn{1}{c}{16} & \multicolumn{1}{c}{Cyan}        & \multicolumn{1}{c}{Ge(4s,4p$_x$,4p$_y$,4p$_z$) - O(2p$_x$,2p$_y$,2p$_z$)}     \\ 
		\multicolumn{1}{c}{17.2} & \multicolumn{1}{c}{17.2}   & \multicolumn{1}{c}{Blue-violet}  & \multicolumn{1}{c}{Ge(4s,4p$_x$,4p$_y$,4p$_z$) - O(2p$_x$,2p$_y$,2p$_z$)}     \\ 
		\multicolumn{1}{c}{20.3} & \multicolumn{1}{c}{20.3} & \multicolumn{1}{c}{Light-green}      & \multicolumn{1}{c}{Ge(4s,4p$_x$,4p$_y$,4p$_z$) - O(2p$_x$,2p$_y$,2p$_z$)}     \\ 
		\multicolumn{1}{c}{24.5} & \multicolumn{1}{c}{24.5} & \multicolumn{1}{c}{Dark-gray} & \multicolumn{1}{c}{Ge(4s) - O(2s)}   \\
		\multicolumn{1}{c}{27.1} & \multicolumn{1}{c}{27.1} & \multicolumn{1}{c}{Gray}      & \multicolumn{1}{c}{Ge(4s) - O(2s)}   \\ \hline\hline
			                                 
	\end{tabular}  
\end{table*}
\hspace*{1 cm}As for $\epsilon_1(\omega)$ (Fig. 5(a)), their prominent absorption structures can be understood from those of $\epsilon_2(\omega)$ by the Kramers-Kronig relation under the event of optical excitations. Apparently, the van Hove singularities in $\epsilon_1(\omega)$ and $\epsilon_2(\omega)$ might be similar or different, being strongly co-related by the principal-value integration on the complex plane. Below the threshold frequency, $\epsilon_1(\omega)$ weakly depends on $\omega$, in which its value roughly lies within the range of 2.1-2.2 for $x$-, $y$-polarizations and the $z$ one. This will determine the low-frequency reflectance spectrum (Fig. 6(b)) and the vanishing absorption coefficient (Fig. 6(c)). Very interesting, $\epsilon_1(\omega)$ is very sensitive to the changes of frequency during the creation of the excited holes and electrons. It can vanish at weak Landau damping, in which its zero point and the small $\epsilon_2(\omega)$ might appear simultaneously, e.g., the vanishing $\epsilon_2(\omega)$ at 18.5/20/13.6 eV under \textbf{E}$\|\hat{\textbf{x}}$/\textbf{E}$\|\hat{\textbf{y}}$/\textbf{E}$\|\hat{\textbf{z}}$. In addition, the zero points for the $z$-polarization (the blue curve) at 6 eV become meaningless because of the combination with very strong electron-hole excitations.\\
\hspace*{1 cm}The energy loss function (ELF), being defined as Im$\left[\frac{-1}{\epsilon(\omega)}\right]$\cite{31}, is the screened response function due to the significant valence charges of the Li, Ge and O atoms. The charge screenings can determine the coherent carrier oscillations at the long wave-length limit during the optical transitions. In general, the collective excitations (a plasmon mode) are revealed as a sufficiently strong peak with ELF higher than 1, as clearly indicated in Fig. 6(a). The strongest peak comes into existence at $\omega_p$=18.5/20/13.6 eV for the $x$-/$y$-/$z$-direction electric polarizations, being attributed to the significant Li-2s, O-(2p$_x$, 2p$_y$, 2p$_z$) $\&$ Ge-(4s, 4p$_x$, 4p$_y$, 4p$_z$) orbitals. The O-2s contribution for the most prominent plasma wave is ignored, since its dominating valence states only appear below -22 eV (the black curve in Fig. 4(d)). In addition, a few minor plasmon peaks are accompanied by serious Landau dampings. It should be noted that two manners could be utilized to identify the collective excitations, but the current peak in ELF is much better than the zero points in $\epsilon_1(\omega)$. The absence of the latter and/or the combination with a very large $\epsilon_2(\omega)$  is the main reason.\\
\hspace*{1 cm}The reflectance spectrum is characterized as R($\omega$)=$\left|\frac{\sqrt{\epsilon(\omega)}-1}{\sqrt{\epsilon(\omega)}+1}\right|^2$ \cite{32}, when an electromagnetic wave is normally incident on Li$_2$GeO$_3$. The total electric field will be reflected by the surface, absorbed by the valence electrons and transmitted through a finite-width sample. Both the reflectance R($\omega)$ and the absorption coefficient $\alpha(\omega)$ are inverse to the propagation decay length in [33], as shown in Figs. 6(b) and 6(c), respectively, directly reflecting the main features of the single-particle (Figs. 5(a) and 5(b)) and the collective excitations (Fig. 6(a)). For $\omega <$ E$_g^o$, a small reflectance weakly depends on the frequency and is roughly given by$\left|\frac{\sqrt{\epsilon(0)}-1}{\sqrt{\epsilon(0)}+1}\right|^2$. The absorption coefficient $\alpha(\omega)=0$  because of the vanishing electron-hole excitations, leading to the non-decay EM-wave propagation and thus a very efficient transmission. However, within the active region of valence-electron excitations (E$_g^o < \omega <$ 21 eV), the sensitive and significant frequency dependences come into existence. Reflectance is enhanced and displays a large fluctuation under the various inter-band excitation channels, in which a drastic change of the plasmon edge appears at $\omega_p$, e.g., $\sim$ 40 \% variation under the \textbf{E}$\|\hat{\textbf{y}}$ case (the red curve). Moreover, since absorption coefficient $\alpha(\omega)$ is very large, especially that related to the plasmon mode, and its inverse is about 50-300 \AA, the EM waves are easily absorbed by Li$_2$GeO$_3$ through the rich electronic excitations.\\
\begin{figure}[t]
	\includegraphics[scale=0.9]{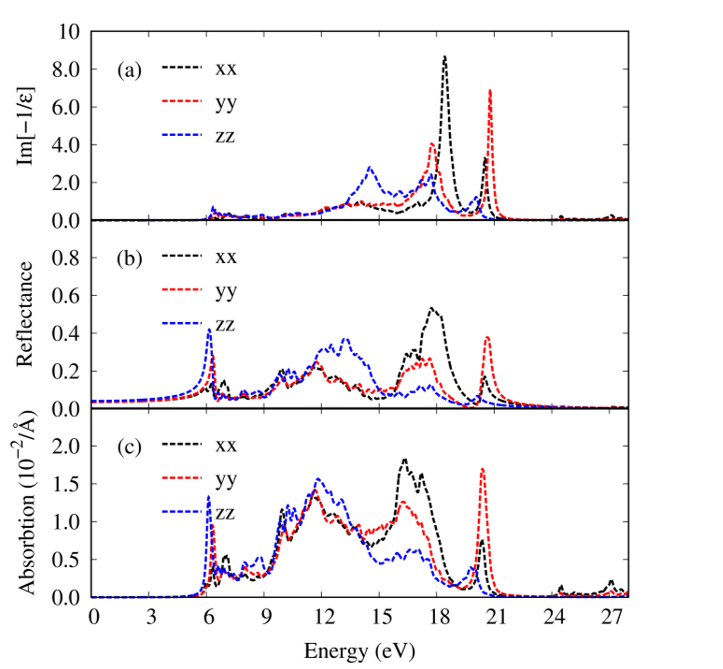}
	\label{fgr:5}
	\caption{The various optical properties: (a) energy loss functions, (b) reflectance spectra, and (c) absorption coefficients.}
\end{figure}
\hspace*{1cm}In addition to X-ray diffractions \cite{12,29}, only very few experimental examinations on electronic and optical properties exist. In general, the wave-vector dependences of occupied valence states can be directly tested by angle-resolved photoemission spectroscopy (ARPES) \cite{16,34}. A lot of unusual energy subbands (Fig. 2) due to a {$Moir\acute{e}$} superlattice would create high barriers in the ARPES measurements. Scanning tunneling spectroscopy (STS) is available for the clear identification of van Hove singularities near the Fermi level \cite{35}. However, STS might be suitable only under a thin-film sample because of very weak quantum currents. Very interestingly, the optical spectroscopy methods of reflectance \cite{17}, absorption \cite{18} and transmission \cite{19} are reliable in verifying the frequency-dependent optical properties. For example, the measured reflectance spectrum, being supported by the Kramers-Kronig relations between $\epsilon_1(\omega)$ and $\epsilon_2(\omega)$, can determine both of them. It provides significant information about the initial excitonic peaks, the greatly reduced threshold excitation frequency, many prominent absorption structures, and the strongest plasmon mode at $\omega_p >$ 16 eV. Up to now, one photo-luminescence measurement has verified the optical gap situated at $\sim$ 6 eV \cite{12}, being close to the current prediction of $\sim$ 5.95 eV. Apparently, the diverse optical properties in cathode/electrolyte/anode materials of Li$^+$-based batteries are worthy of systematic investigations, both experimentally and theoretically.\\
\section{CONCLUSION}
\hspace*{1cm} The significant orbital hybridizations in chemical bonds, being based on first-principles calculations, are thoroughly identified for the Li$_2$GeO$_3$ compound in terms of the geometric, electronic and optical properties. They will play important roles in fully understanding the diversified essential properties of cathode/electrolyte/anode materials in ion-related materials. This solid-state electrolyte presents unusual features, i.e. a {$Moir\acute{e}$} superlattice with a highly non-uniform chemical environment, the Li-, Ge- or O-dominated energy bands, orbital-induced spatial charge densities, atom- and orbital-decomposed van Hove singularities. As a result, the band-edge states, which might create the prominent optical responses, are well characterized by the specific orbital interactions through the developed theoretical framework. The featured optical transitions cover a red-shift optical gap (E$_g^o$ = 5.95 eV) much lower than an indirect one (E$_g^i$ = 6.9 eV), a very long transmission length/low reflectance for $\omega <$ E$_g^o$, 16 pronounced single-particle absorption structures (short decay lengths), and the strongest plasmon peak/a quickly decreasing edge in the energy loss function/reflectance spectrum at $\omega_p$, 17-18 eV, and sensitive changes due to the electric-field directions. The many-body excitonic effects have strongly modified the single-particle interband excitations. Under current investigations, the developed theoretical framework can be further generalized to include other electrolyte/cathode/anode materials of Li$^+$-based batteries.
\section{Acknowledgments}
 \hspace*{1cm}This work is supported by the Hi-GEM Research Center and the
 Taiwan Ministry of Science and Technology under grant number MOST 108-2212-M-006-022-MY3, MOST 109-2811-M-006-505 and MOST 108-3017-F-006-003.

\end{document}